\documentclass[prc,twocolumn,showpacs,superscriptaddress,preprintnumbers,amsmath,amssymb,nofootinbib]{revtex4-1}
\usepackage{bm}
\usepackage{dcolumn}
\usepackage{multirow}
\usepackage{ifpdf}
\usepackage{url}

\ifpdf
\usepackage{graphicx} 
\usepackage{hyperref}
\else 
\usepackage[dvipdfmx]{graphicx} 
\usepackage[dvipdfmx]{hyperref}
\fi
\usepackage{color} 

\usepackage{slashed}
\usepackage{bm}
\usepackage{dcolumn}
\usepackage{multirow}
\usepackage{ifpdf}
\usepackage{url}
\usepackage{here}
\usepackage{graphicx}
\usepackage{indentfirst}
\usepackage{listings}

\usepackage{ifthen}
\usepackage{kpfonts}
\usepackage{tgtermes}
\usepackage{amsmath,amssymb}
\usepackage{fancyhdr}
\usepackage{titletoc}
\usepackage{xcolor}

\usepackage{mathtools}

\newcommand{\dndeta}[1][flat]
{
    \ifthenelse{\equal{#1}{flat}}{$\left< dN_{\mathrm{ch}}/d\eta\right>$}{}
    \ifthenelse{\equal{#1}{vertical}}{$\left< \cfrac{dN_{\mathrm{ch}}}{d\eta\right>}$}{}
}

\newcommand{\snn}[1][nucleus]
{
    \ifthenelse{\equal{#1}{nucleus}}{$\sqrt{s_{\mathrm{NN}}}$}{}
    \ifthenelse{\equal{#1}{proton}}{$\sqrt{s}$}{}
}

\newcommand{\ctwofour}{c_{2}\{4\}}

\newcommand\pythia{\textsc{Pythia}}
\newcommand\ISthreeD{\textsc{iS3D}}
\newcommand\jam{\textsc{Jam}}

\hypersetup{
colorlinks=true,
linkcolor=blue,
citecolor=blue,
urlcolor=blue
}

\begin{document}

\title{
Non-equilibrium components in very low transverse momentum region\\
in high-energy nuclear collisions
}
\author{Yuuka Kanakubo}
\email{y-kanakubo-75t@eagle.sophia.ac.jp}
\affiliation{%
Department of Physics, Sophia University, Tokyo 102-8554, Japan
}

\author{Yasuki Tachibana}
\email{ytachibana@aiu.ac.jp}
\affiliation{
Akita International University, Yuwa, Akita-city 010-1292, Japan
}
\author{Tetsufumi Hirano}
\email{hirano@sophia.ac.jp}
\affiliation{%
Department of Physics, Sophia University, Tokyo 102-8554, Japan
}

\date{\today}

\begin{abstract}
We analyze Pb+Pb collisions at $\sqrt{s_{\mathrm{NN}}}=2.76$ TeV with a novel framework based on the dynamical core--corona picture that describes particle productions from both equilibrium and non-equilibrium components.
We remark the possibility of the contribution from non-equilibrium components at very low transverse momentum ($p_{T}$) region
and show that such contributions significantly affect $p_{T}$-integrated four-particle cumulants.
These results strongly suggest the necessity of non-equilibrium components when one extracts properties of the quark-gluon plasma from experimental data using sophisticated dynamical models based on relativistic hydrodynamics.

\end{abstract}

\pacs{25.75.-q, 12.38.Mh, 25.75.Ld, 24.10.Nz}

\maketitle

\textit{Introduction.} Properties of the quark-gluon plasma (QGP) \cite{Collins:1974ky,Cabibbo:1975ig,Shuryak:1977ut}, 
the primordial matter of our universe, have been investigated through the relativistic heavy-ion collision experiments at the Relativistic Heavy Ion Collider (RHIC) and Large Hadron Collider (LHC).
Since the discovery of the nearly perfect fluidity of the QGP
solidified the applicability of relativistic hydrodynamics \cite{Heinz:2001xi,Gyulassy:2004zy,Shuryak:2004cy,Hirano:2005wx},
one of the main streams of the theoretical QGP study has been 
led by multi-stage dynamical models in which relativistic hydrodynamics plays a central role in describing the dynamics of the QGP.
Since then, there have been several splendid theoretical developments on the dynamical models.

The particle spectra at low transverse momentum region ($p_{T} \lesssim 5$ GeV) 
are commonly understood as thermal distributions boosted by the radial velocity of expansion which can be described by theoretical descriptions based on relativistic hydrodynamics: dynamical models with the hydrodynamic description of the QGP
(for a review, see, e.g., Ref.~\cite{Shen:2021nbe})
or the blast wave model \cite{Schnedermann:1993ws}.
However, contrary to the common understanding, 
it has also been widely discussed that there is a significant lack of particle production in hydrodynamic models 
at the very low $p_{T}$ region ($p_{T} \lesssim 1$ GeV/$c$) compared to experimental data.
Back in the late 90s, the possible interpretations of the discrepancy at the very low $p_T$ were suggested:
the existence of pion condensation \cite{Zimanyi:1979ga}, effects of non-chemical equilibrium \cite{Kataja:1990tp,Gavin:1991ki},
and contribution from resonance decays with the blast wave model \cite{Sollfrank:1991xm} and the hydrodynamic simulation \cite{Ornik:1991dm}.
Although it seemed that the issue is almost resolved by the above works at that time,
it has been still reported that the state-of-the-art hydrodynamic models cannot describe the particle yields at the very low $p_T$ at the LHC energies even if Bayesian parameter estimation or global fit are performed \cite{Nijs:2020ors,Devetak:2019lsk}.
Thus, to our best knowledge, no successful interpretation has been made to the spectra at the very low $p_{T}$ region
while several interpretations are suggested as possible solutions \cite{Begun:2013nga,Begun:2014rsa,Begun:2015ifa,Huovinen:2016xxq,Guillen:2020nul,Grossi:2021gqi}.
Therefore, other promising mechanisms are required to explain the excess of particle yields in experimental data over conventional hydrodynamic models.

\begin{figure*}[htpb]
    \centering
    \includegraphics[bb=0 0 628 538, width=0.45\textwidth]{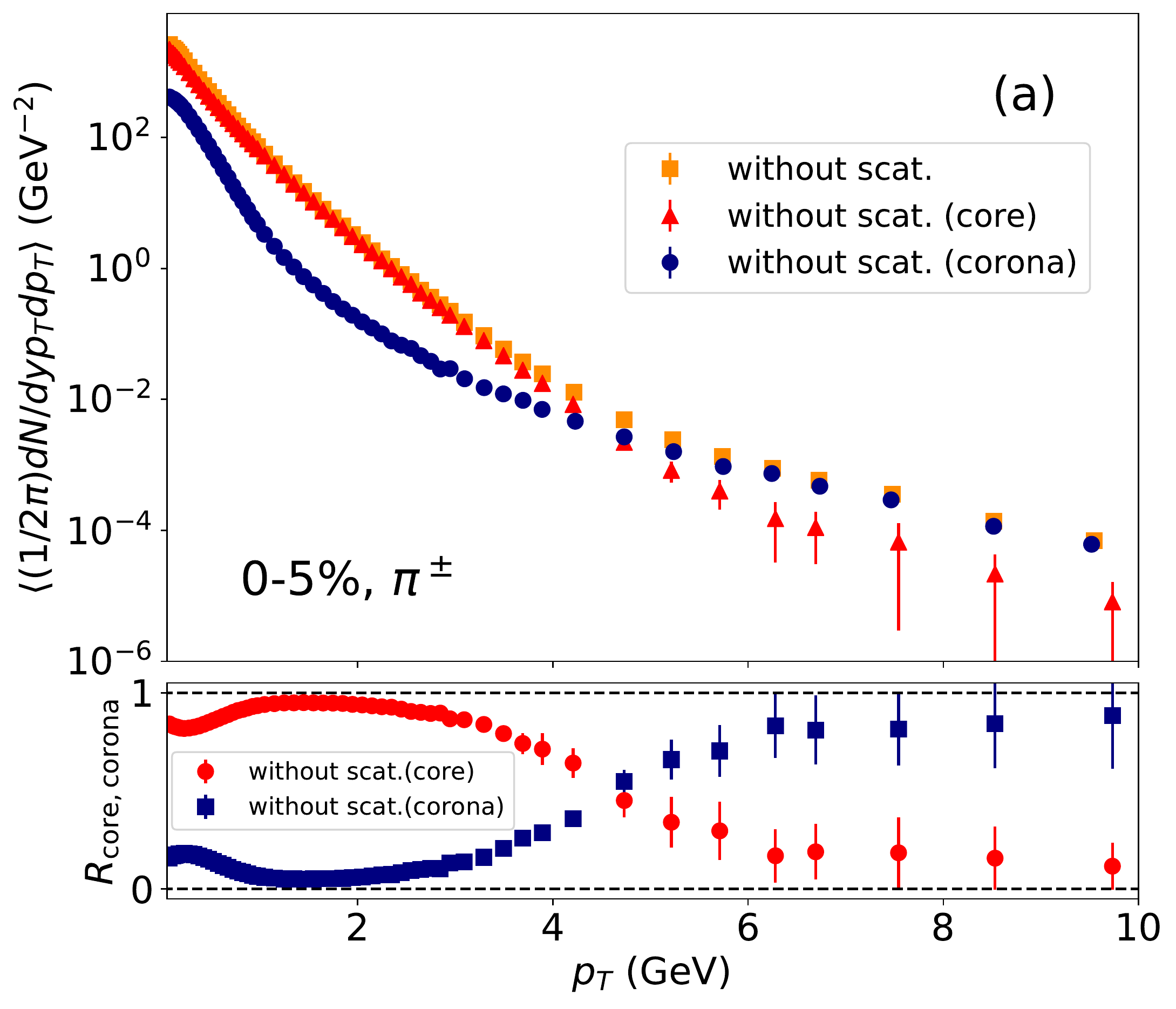}
    \vspace{13pt}
    \includegraphics[bb=0 0 628 538, width=0.45\textwidth]{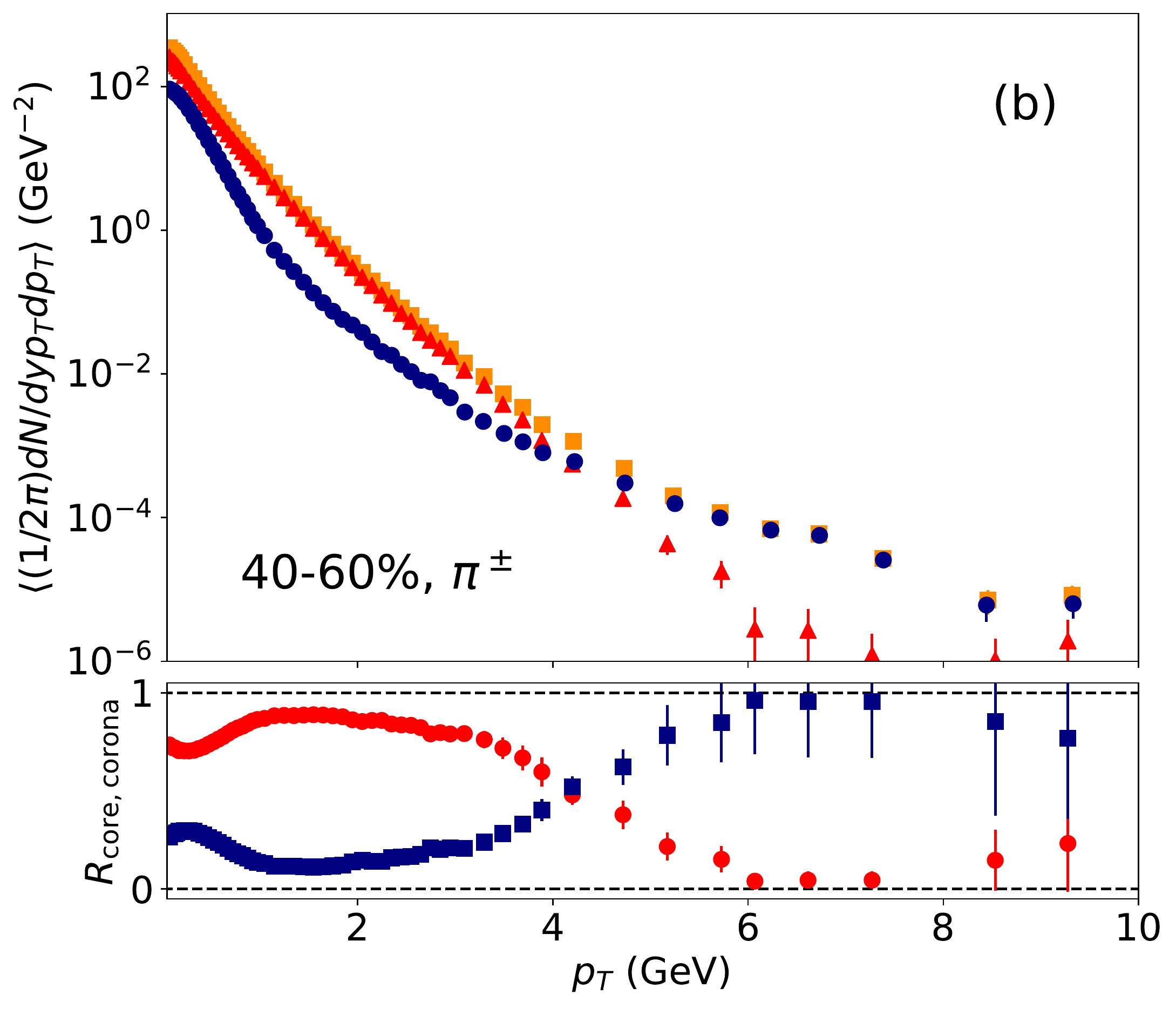}
    \vspace{13pt}
    \includegraphics[bb=0 0 628 538, width=0.45\textwidth]{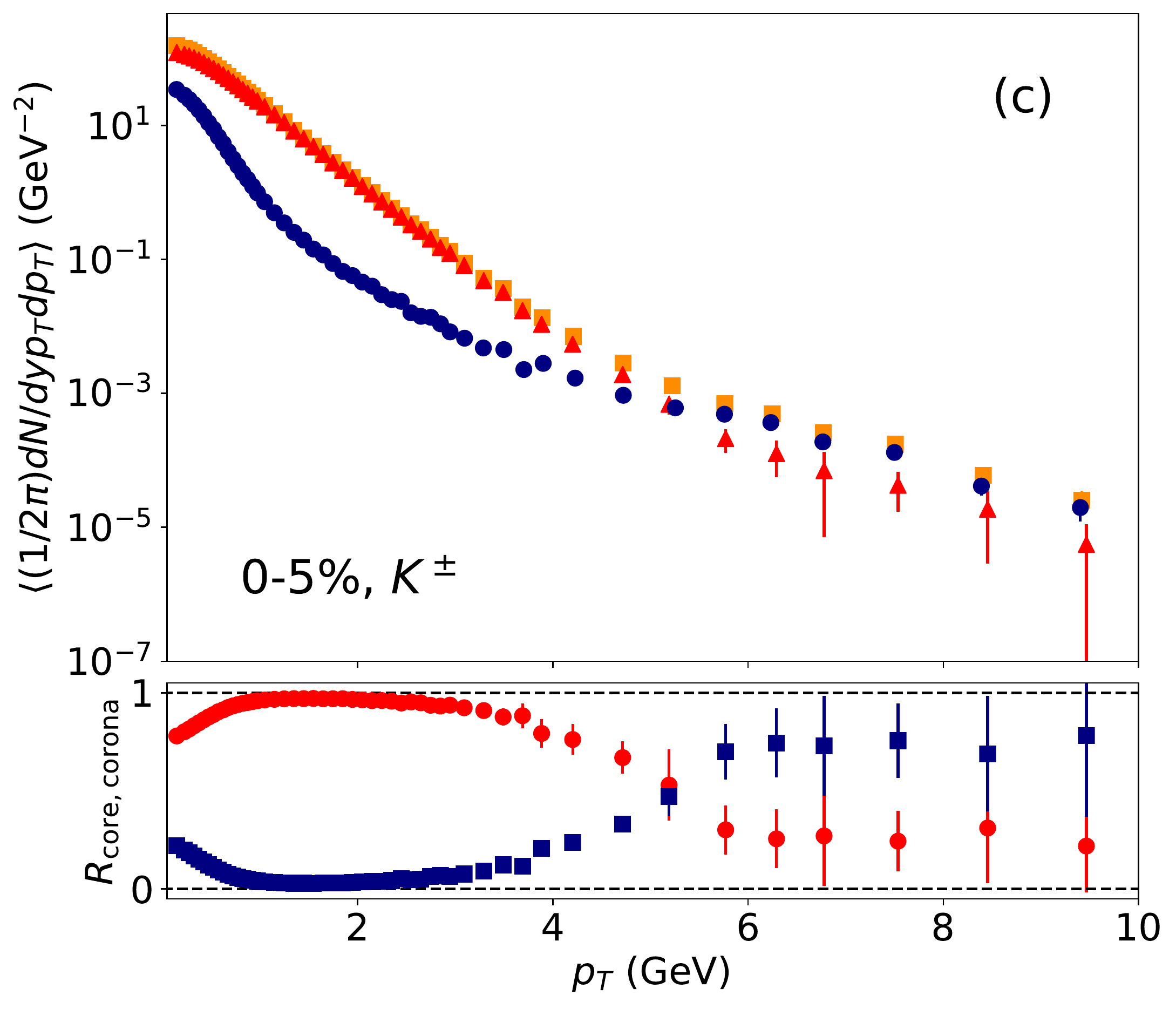}
    \vspace{13pt}
    \includegraphics[bb=0 0 628 538, width=0.45\textwidth]{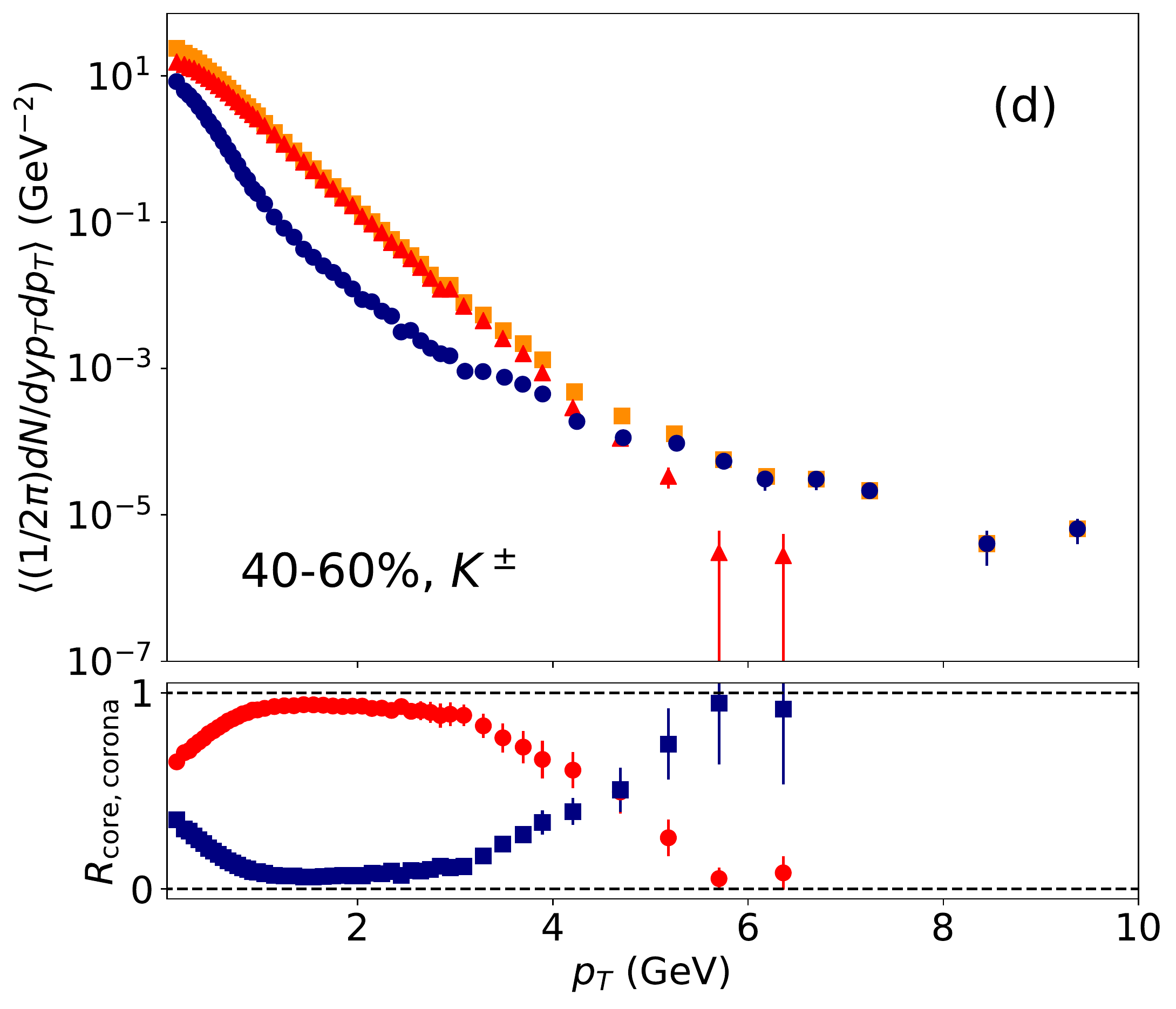}
    \vspace{13pt}
    \includegraphics[bb=0 0 628 538, width=0.45\textwidth]{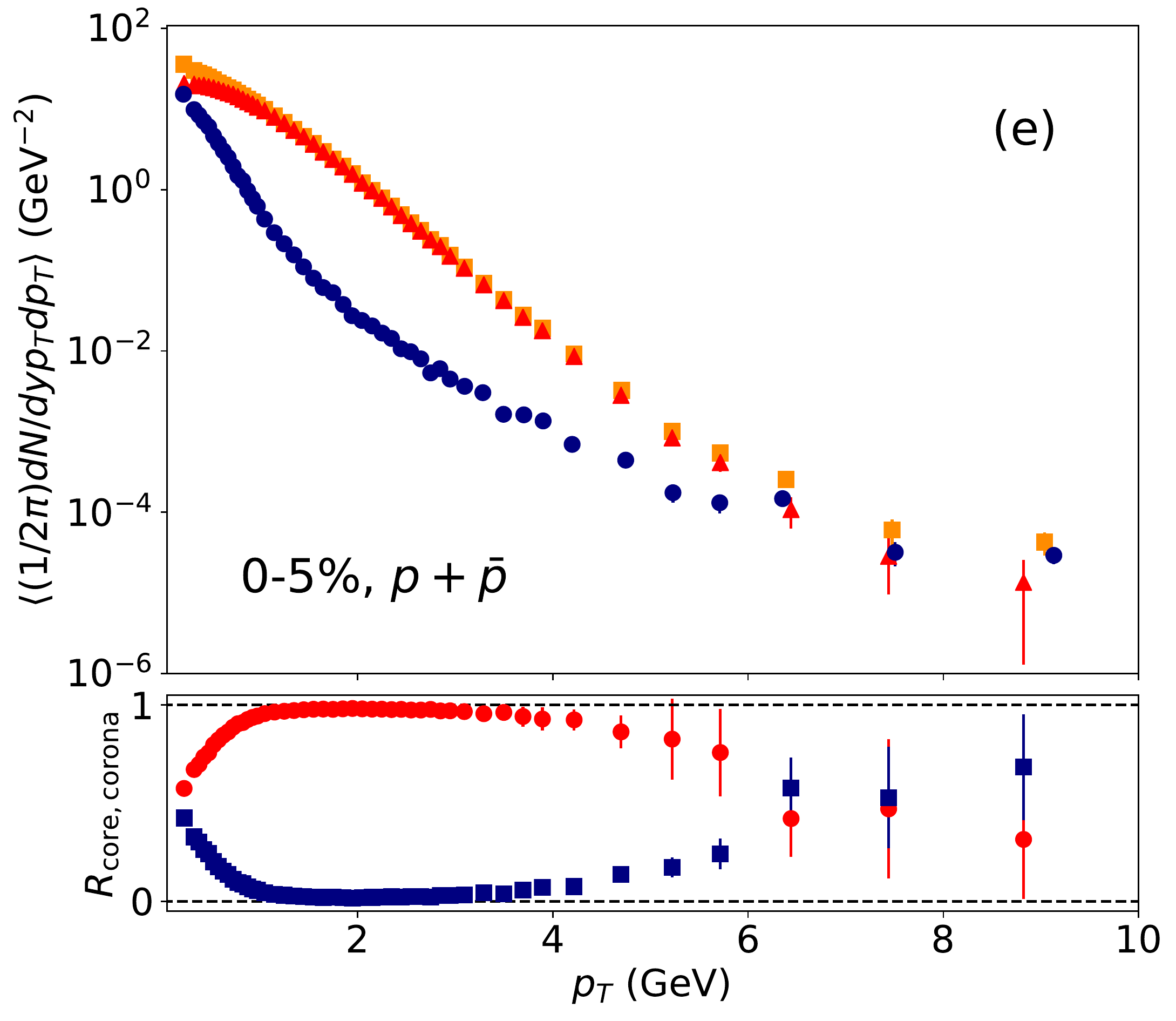}
    \vspace{13pt}
    \includegraphics[bb=0 0 628 538, width=0.45\textwidth]{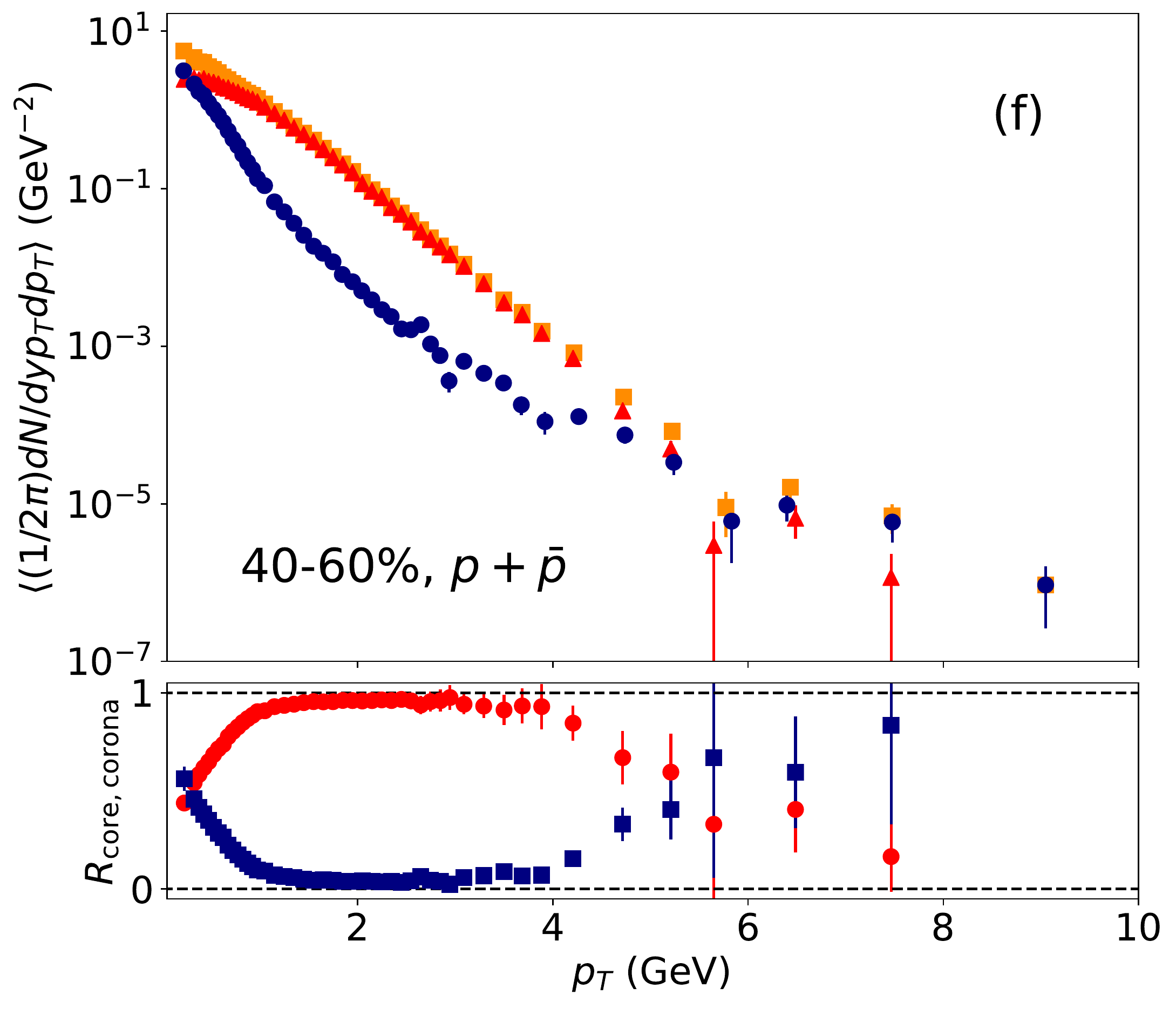}
    \caption{Transverse momentum spectra of charged pions in (a) 0-5\% and (b) 40-60\%, charged kaons in (c) 0-5\% and (d) 40-60\%, and protons and antiprotons in (e) 0-5\% and (f) 40-60\% centrality classes, in Pb+Pb collisions at \snn = 2.76 TeV from DCCI2.
    (Upper) Results with switching off hadronic rescatterings (orange squares) and their breakdown into core (red triangles) and corona contributions (blue circles) are shown.
    (Lower) Fraction of the core $R_{\mathrm{core}}$ (red circles) and that of the corona $R_{\mathrm{corona}}$ (blue squares) components in each $p_{T}$ bin.}
    \label{fig:CORECORONA_PTSPECTRA}
\end{figure*}

In this Letter, we investigate this problem with the dynamical core--corona initialization framework (DCCI2) \cite{Kanakubo:2021qcw}.
This framework was built aiming for a comprehensive description of high-energy nuclear collisions from small to large colliding systems, {\textit{i.e.,}} proton--proton ({\textit{p}}--{\textit{p}}) to heavy-ion (A--A) collisions.
In this framework, 
dynamical aspects of the core--corona picture \cite{Werner:2005jf} are attained by the novel dynamical initialization framework \cite{Okai:2017ofp,Shen:2017bsr}. 
Under the core--corona picture \cite{Werner:2005jf}, the systems generated in high-energy nuclear collisions are described with two components: 
the core (equilibrated matter) and the corona (non-equilibrated matter).
The hadronic production from the core components exhibits the boosted equilibrium distribution, 
while the corona components undergo string fragmentation in a vacuum. 
In our previous work \cite{Kanakubo:2021qcw}, we found a non-trivial interplay between the core and the corona components in $p_{T}$ spectra of charged hadrons in minimum bias Pb+Pb collisions.
As expected, 
the core components are dominant in $p_{T} \lesssim 4$ GeV/$c$ due to the QGP fluid formation, while the corona components, which are the leading hadrons originating from the hard partons, are dominant in high $p_{T}$ regions.
Surprisingly, it turns out that the corona components reach $\approx 20$\%  below $p_{T} \approx 1$ GeV. These soft hadrons are fragmented from strings consisting of hard partons.
Such a non-negligible contribution to the particle productions from corona components at midrapidity gives a certain correction to observables obtained purely from core components described by hydrodynamics.

In this Letter, using DCCI2, we analyze the ``soft from corona" components in particle-identified hadron $p_{T}$ spectra and the centrality dependence of $p_{T}$ spectra of charged pions and elucidate the effects of the corona components on the four-particle cumulant, $c_{2}\{4\}$, in Pb+Pb collisions at \snn = 2.76 TeV.
Throughout this Letter, 
we use the natural unit, $\hbar = c = k_{B} =1$.


\textit{Model.} The dynamical separation into the core and corona components is realized by implementing the core--corona picture into the dynamical initialization framework \cite{Kanakubo:2018vkl,Kanakubo:2019ogh,Kanakubo:2021qcw}.
Under this framework, initial conditions of relativistic hydrodynamics are dynamically generated via source terms of hydrodynamic equations.
With the assumption that the system is described with the QGP fluids (the core) and the non-equilibrated partons (the corona) and that the energy and momentum are conserved as a summation of the two components,
energy and momentum deposited from non-equilibrated partons become the sources of QGP fluids.

The entire flow of DCCI2 \cite{Kanakubo:2021qcw} is summarized as follows:
The system produced at the initial state is described by phase-space distributions of initially produced partons obtained from \pythia8.244 \cite{Sjostrand:2007gs} with switching off hadronization on an event-by-event basis.
The heavy-ion mode of \pythia, \pythia8 Angantyr \cite{Bierlich:2016smv,Bierlich:2018xfw}, is used for simulations of heavy-ion collisions.
At a formation time, $\tau_0 = 0.1$ fm, of initially generated partons, dynamical initialization of the QGP fluids based on the core--corona picture starts and the system is getting separated into the core and corona components.
For the space-time evolution of the core, ideal hydrodynamic simulations are performed in the (3+1)-dimensional Milne coordinates \cite{Tachibana:2014lja} incorporating the $s$95$p$-v1.1 \cite{Huovinen:2009yb}, the equation of state constructed with a seamless connection of the (2+1)-flavor lattice QCD at high temperature and a hadron resonance gas model at low temperature.
The particlization of fluids is performed with \ISthreeD \ \cite{McNelis:2019auj}, a Monte-Carlo sampler converting hydrodynamic fields on the hypersurface at the switching temperature, $T_{\mathrm{sw}}=0.165$ GeV, into hadrons.
The non-equilibrated partons undergo hadronization through the Lund string fragmentation with \pythia8.
The hadrons obtained from both switching hyper-surface and string fragmentation are handed to the hadronic transport model, \jam \ \cite{Nara:1999dz}, to perform hadronic rescatterings and resonance decays.

In the following, we denote ``full DCCI2 simulations" by all steps mentioned above.
To see a breakdown of the total yields into the core and the corona components, we switch off hadronic rescatterings in \jam\ and perform only resonance decays.
This is because hadronic rescatterings mix up the two components in the late stage of reactions.
We also have an option ``the core components with hadronic rescatterings" in which we neglect the corona components from string fragmentation in the hadronic cascade and mimics the results from conventional hybrid models in which hydrodynamic evolution is followed by the hadronic afterburner.

Note that we minor-updated the following three parameters of DCCI2 simulations from Ref.~\cite{Kanakubo:2021qcw} to improve description of $p_{T}$ spectra in central Pb+Pb collisions at \snn = 2.76 TeV:
the parameter regulating cross sections of multiparton interactions and infrared QCD emissions in initial parton generation, $p_{\mathrm{T0Ref}}$, the coefficient of the cross section of the collision between two partons in the dynamical core--corona initialization $\sigma_0$, and the transverse width of the Gaussian distribution used in the smearing of deposited energy and momentum in the dynamical generation of the QGP fluids, $\sigma_\perp$.
In this Letter, we use $p_{\mathrm{T0Ref}}$ ($p+p$) = 1.9 GeV,  $p_{\mathrm{T0Ref}}$ (Pb+Pb) = 1.0 GeV, 
$\sigma_0=0.3$ $\mathrm{fm}^2$, and $\sigma_\perp=0.6$ fm, which do not alter the conclusion significantly in Ref.~\cite{Kanakubo:2021qcw}.\footnote{Although we will not discuss results in $p$+$p$ collisions in this Letter, we denote $p_{\mathrm{T0Ref}}$ ($p$+$p$) for the sake of comparison.}
Details of the other parameters can be found in Ref.~\cite{Kanakubo:2021qcw}.

\begin{figure*}[htpb]
    \centering
    \includegraphics[bb=0 0 628 538, width = 0.45 \textwidth]{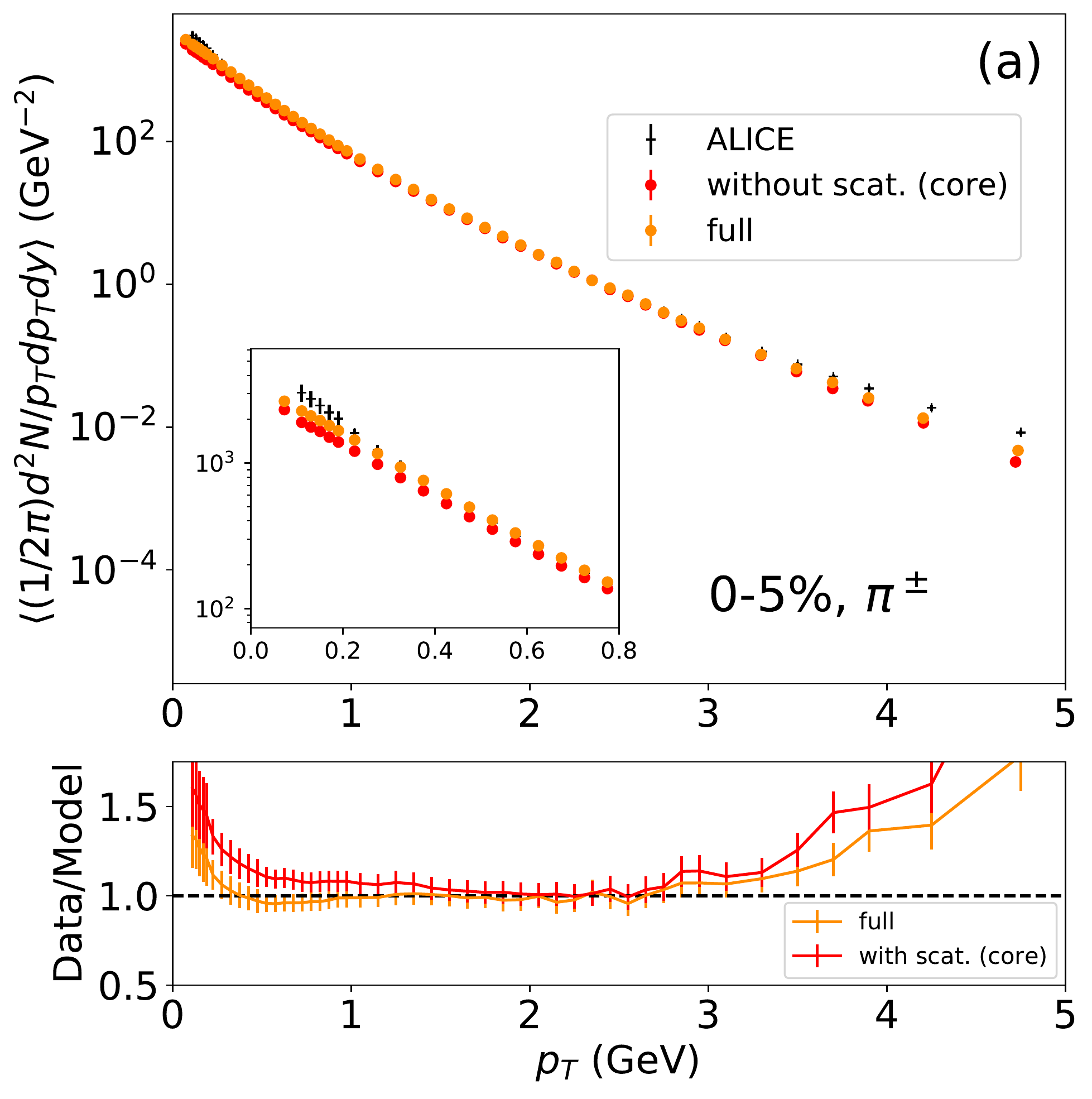}
    \vspace{13pt}
    \includegraphics[bb=0 0 628 538, width = 0.45 \textwidth]{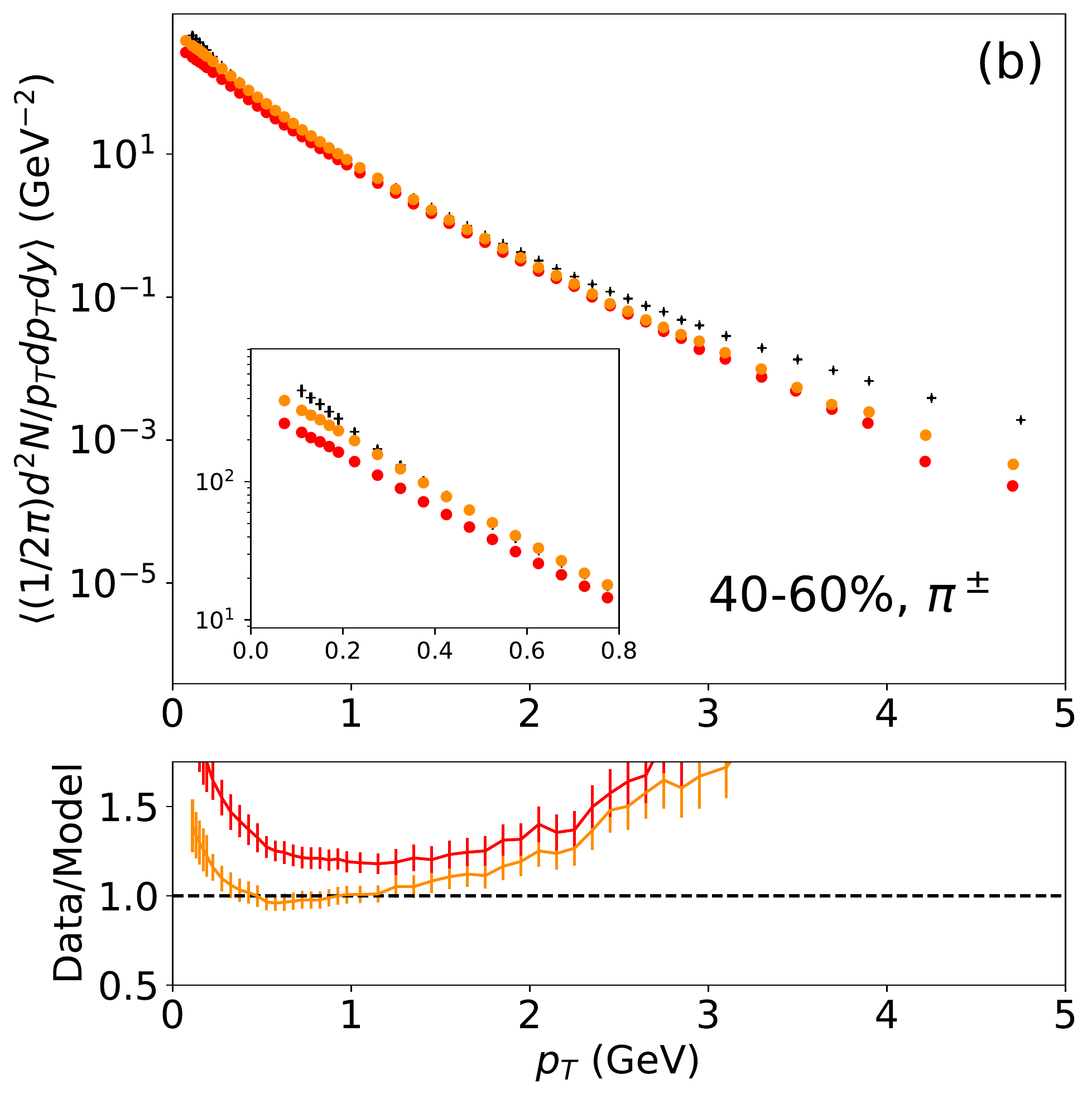}
    \vspace{13pt}
    \includegraphics[bb=0 0 628 538, width = 0.45 \textwidth]{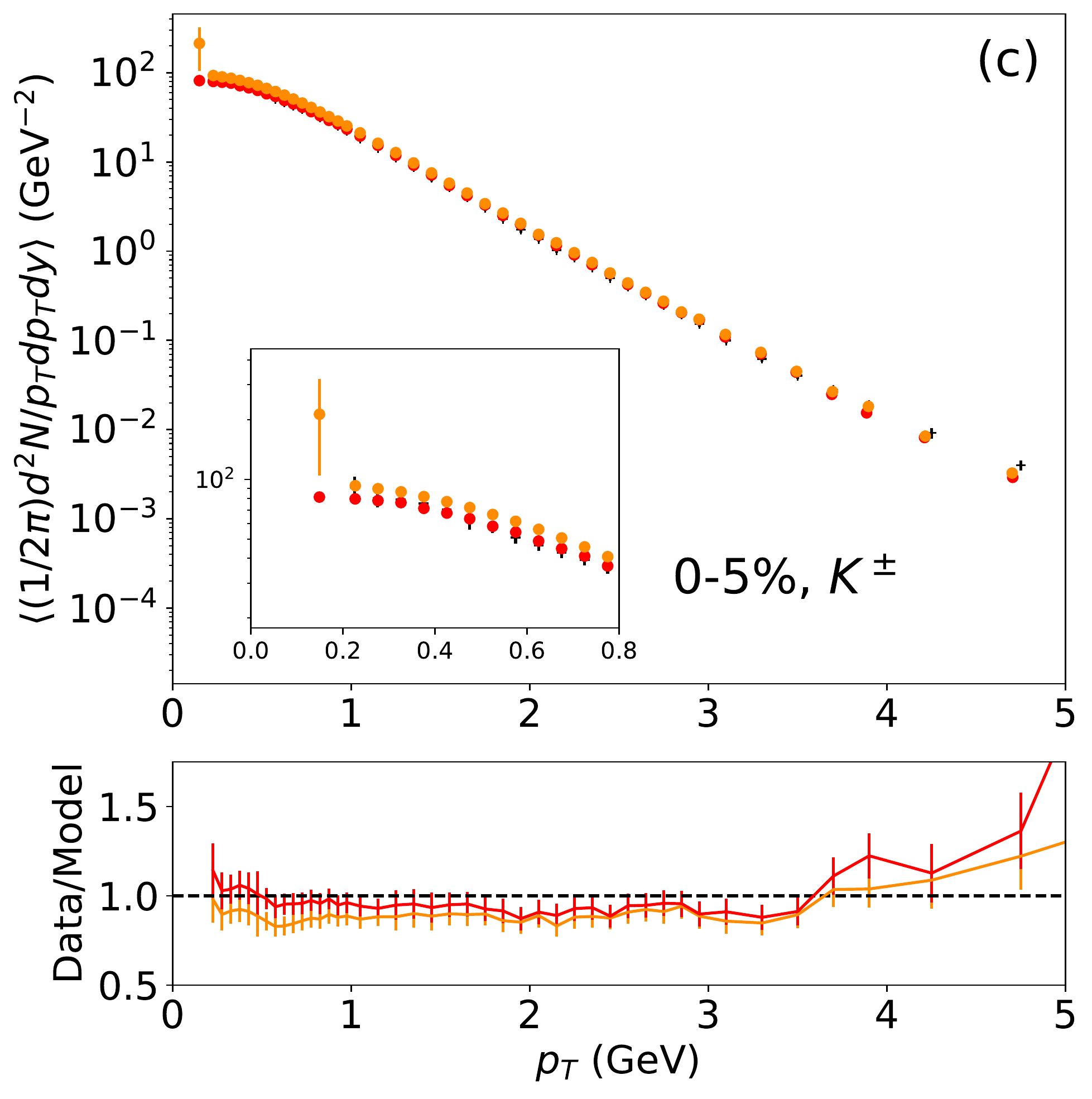}
    \vspace{13pt}
    \includegraphics[bb=0 0 628 538, width = 0.45 \textwidth]{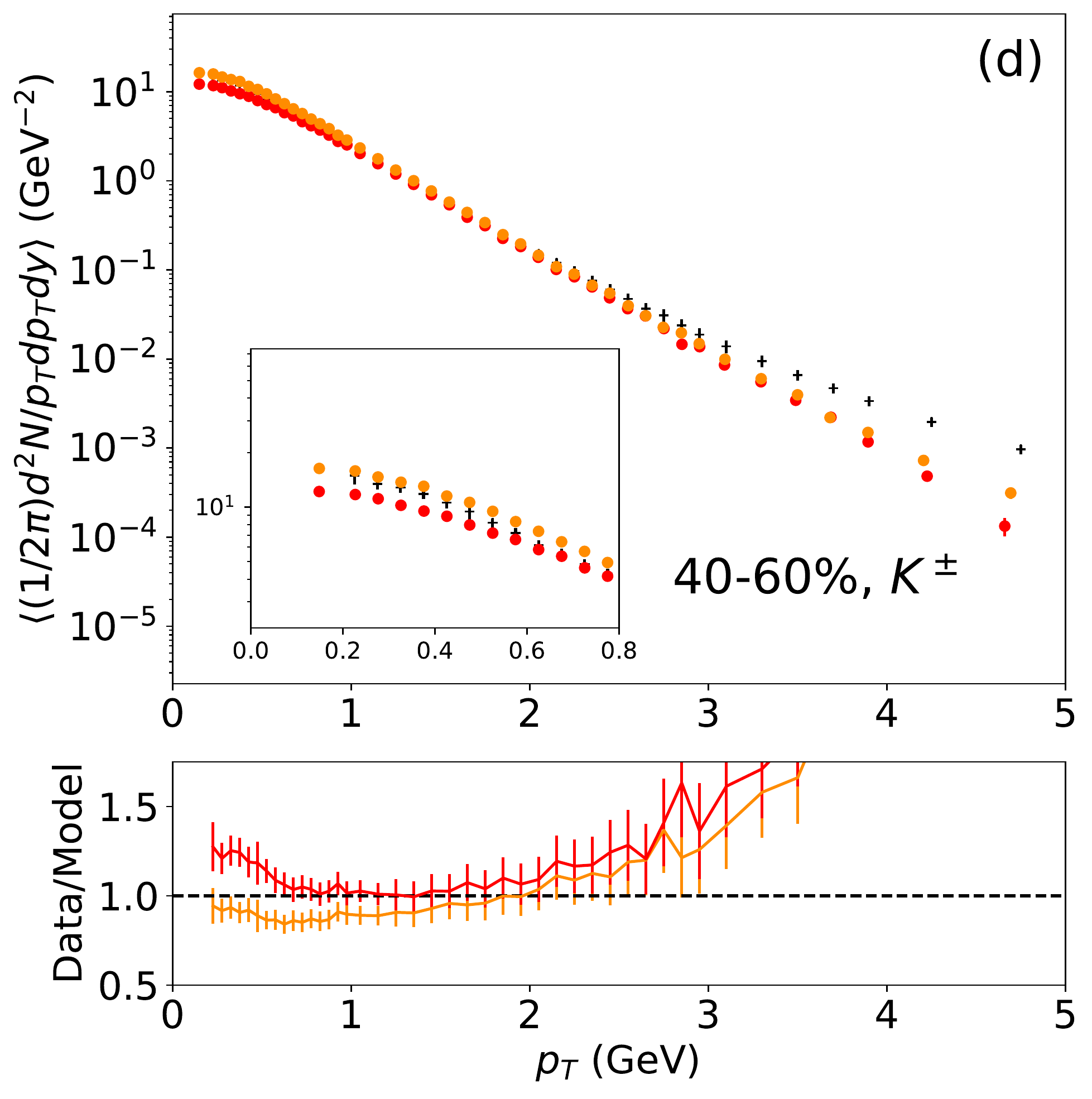}
    \vspace{13pt}
    \includegraphics[bb=0 0 628 538, width = 0.45 \textwidth]{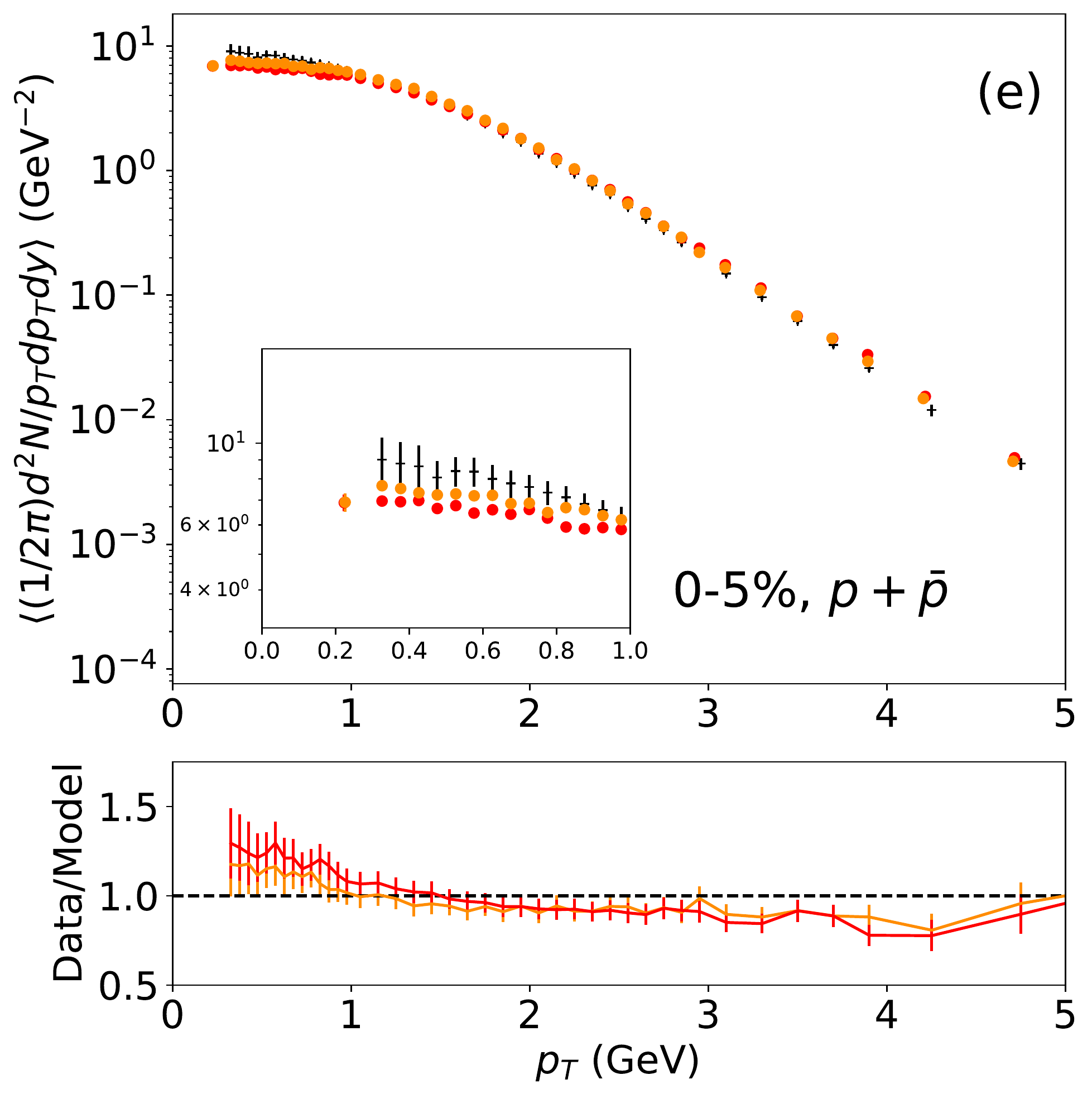}
    \vspace{13pt}
    \includegraphics[bb=0 0 628 538, width = 0.45 \textwidth]{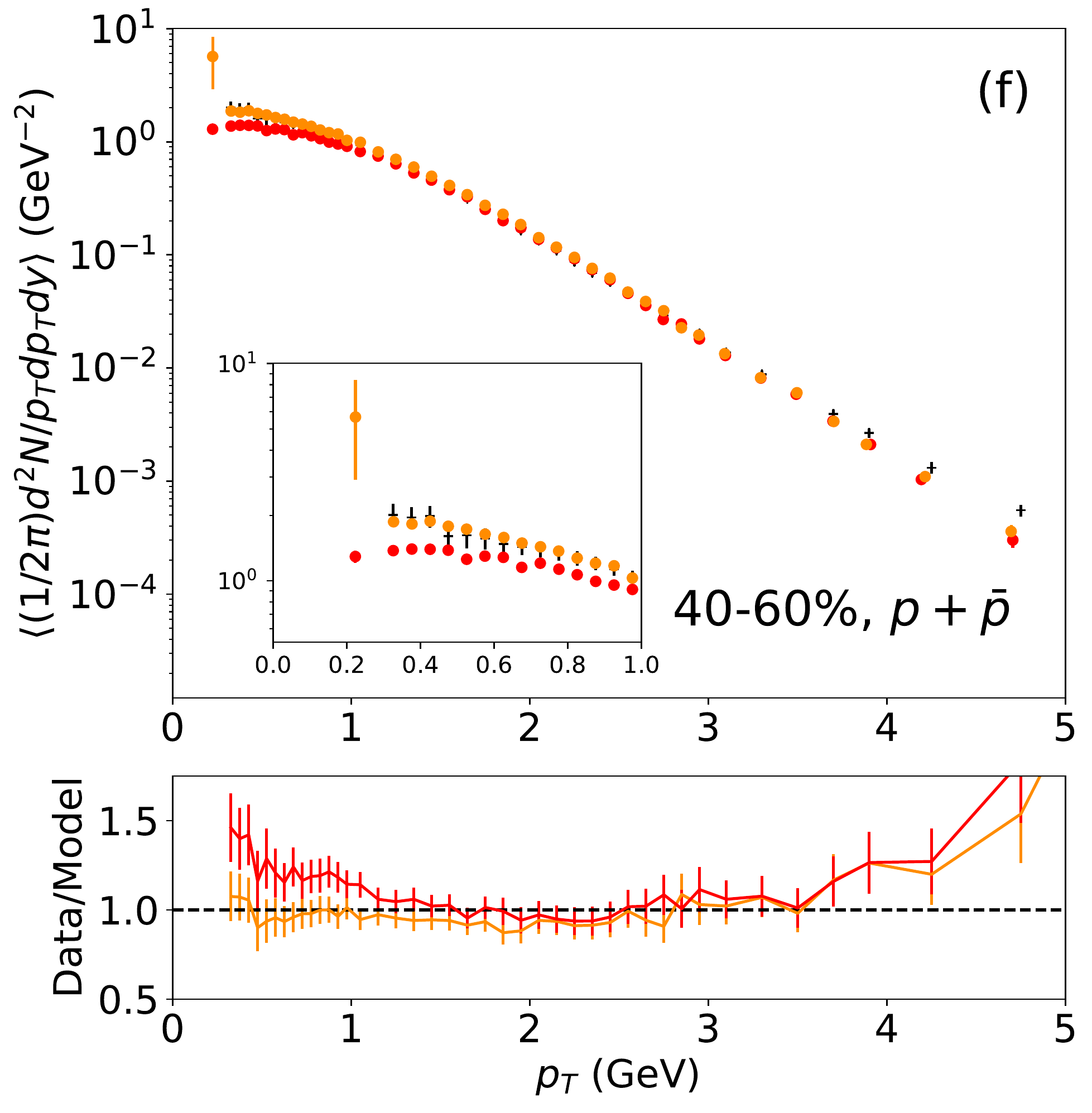}
    \caption{
    Transverse momentum spectra of charged pions in (a) 0-5\% and (b) 40-60\%, charged kaons in (c) 0-5\% and (d) 40-60\%, and protons and antiprotons in (e) 0-5\% and (f) 40-60\% centrality classes, in Pb+Pb collisions at \snn = 2.76 TeV from DCCI2.
    (Upper) Results from full DCCI2 simulations (orange circles) and the core components with hadronic rescatterings (red circles) are compared with ALICE experimental data (black crosses).
    Insets are enlarged plots focusing on very low $p_{T}$ regions.
    (Lower) Ratios of the ALICE experimental data \cite{ALICE:2015dtd} to results from full DCCI2 simulation (orange) and core components with hadronic rescatterings (red) at each $p_{T}$ bin.}
    \label{fig:PTSPECTRA_LOWPT}
\end{figure*}

\textit{Results.} First, we investigate the fraction of core and corona contributions in $p_{T}$ spectra of charged pions, charged kaons, and protons and antiprotons in central and mid-central collisions in Pb+Pb collisions at \snn = 2.76 TeV  to capture the overall tendency.
Second, the comparisons of $p_{T}$ spectra between experimental data and results from DCCI2 are shown
to quantify the contribution of the corona components at the very low $p_{T}$ region.
Finally, we also elucidate the effect of the corona components at the very low $p_{T}$ region on the four-particle cumulant, $c_2\{4\}$, as a function of multiplicity, $N_{\mathrm{ch}}$.

Figure \ref{fig:CORECORONA_PTSPECTRA} shows $p_{T}$ spectra of charged pions, charged kaons, and protons and antiprotons in $|\eta|<0.8$ at 0-5\% and 40-60\% centrality classes in Pb+Pb collisions at \snn = 2.76 TeV.
The hadronic rescatterings are switched off in \jam \ to investigate the contribution from core and corona components.
As an overall tendency for all particle-identified hadrons and centrality, 
the core and corona contribution is dominant in lower and higher $p_{T}$, respectively.
This is exactly the consequence of the implementation of the dynamical core--corona picture: partons with lower momentum tend to deposit energy and momentum and the ones with higher momentum tend to survive in the QGP fluids.

The classification of centrality and particle identification, which were not made in our previous paper \cite{Kanakubo:2021qcw},
reveal further dynamics.
Regarding the core components,
the slopes of $p_{T}$ spectra of core components become slightly flatter in more central events for all hadron species,
which originates from the generation of stronger radial flows in more central events.
The fraction of the core components compared to the corona components is larger in more central events, 
which is also a consequence of the implementation of the core--corona picture.
As a result, the dominance of the core components reaches up to higher $p_{T}$ in more central events for all hadron species.
This tendency is strongly seen in heavier particles such as protons and antiprotons since heavier particles in the core components acquire larger $p_{T}$ due to the mass effect of radial expansion \cite{Hirano:2003pw}.
This leads to the core dominance of protons spanning up to $p_{T}\approx6$ GeV in 0-5\% centrality class.\footnote{The dynamical initialization generates random initial transverse flow due to momentum conservation of initially generated partons \cite{Okai:2017ofp}. 
This results in a broad range of radial flow velocity at the switching hypersurface compared to conventional hydrodynamic simulation with vanishing initial transverse flow. Consequently, the core components are blue-shifted by various radial flow velocities, and the resulting $p_{T}$ slope in the higher $p_{T}$ region tends to reflect the large radial push at the dynamical initialization.
}

Now we focus on the contribution from the corona components in the very low $p_{T}$ region.
The corona contribution is getting more significant for lower $p_{T}$ in $p_{T}\lesssim 1$ GeV, 
which is the same tendency as in Fig.~5 (b) in Ref.~\cite{Kanakubo:2021qcw}.
This behavior can be naively understood as a consequence of the interplay between two different shapes of the spectra, approximately an exponential function $\approx e^{-p_{T}/T}$ ($T>0$) and a power-law function $\approx 
1/p_{T}^{n}$ ($n>0$)  for the core and the corona components, respectively.
Contrary to the exponential spectra that linearly decrease with $p_{T}$ in the semi-logarithmic plot, the slope of the power law spectra increases with decreasing $p_{T}$ and can be steeper than the exponential spectra towards $p_{T}\rightarrow 0$ GeV.
This leads to a rather non-trivial $p_{T}$ dependence of the fraction of the core and the corona components and
the relative increase of the corona contributions compared to the core contributions at such very low $p_{T}$. 
Since the fraction of the corona components increases with decreasing $p_{T}$-integrated yields at midrapidity \cite{Kanakubo:2021qcw},
the more peripheral collisions, the larger contributions from the corona components  are at the very low $p_{T}$ for all hadron species.
Even when we focus on 0-5\% central event, the relative increase of the corona contributions at the very low $p_{T}$ still remains and the fraction of the corona components is even larger in heavier particle spectra.
Noteworthy, the fraction of the corona components of protons and antiprotons reaches $\approx 50\%$ at $p_{T} \rightarrow 0$ GeV at 0-5\% central events.
These tendencies are all understood as a consequence of interplay between the blue shift of the core spectra for heavy particles and the ``soft from corona" components from fragmentation of strings consisting of hard partons.

\begin{figure*}[htpb]
    \centering
    \includegraphics[bb=0 0 634 453, width = 0.45\textwidth]{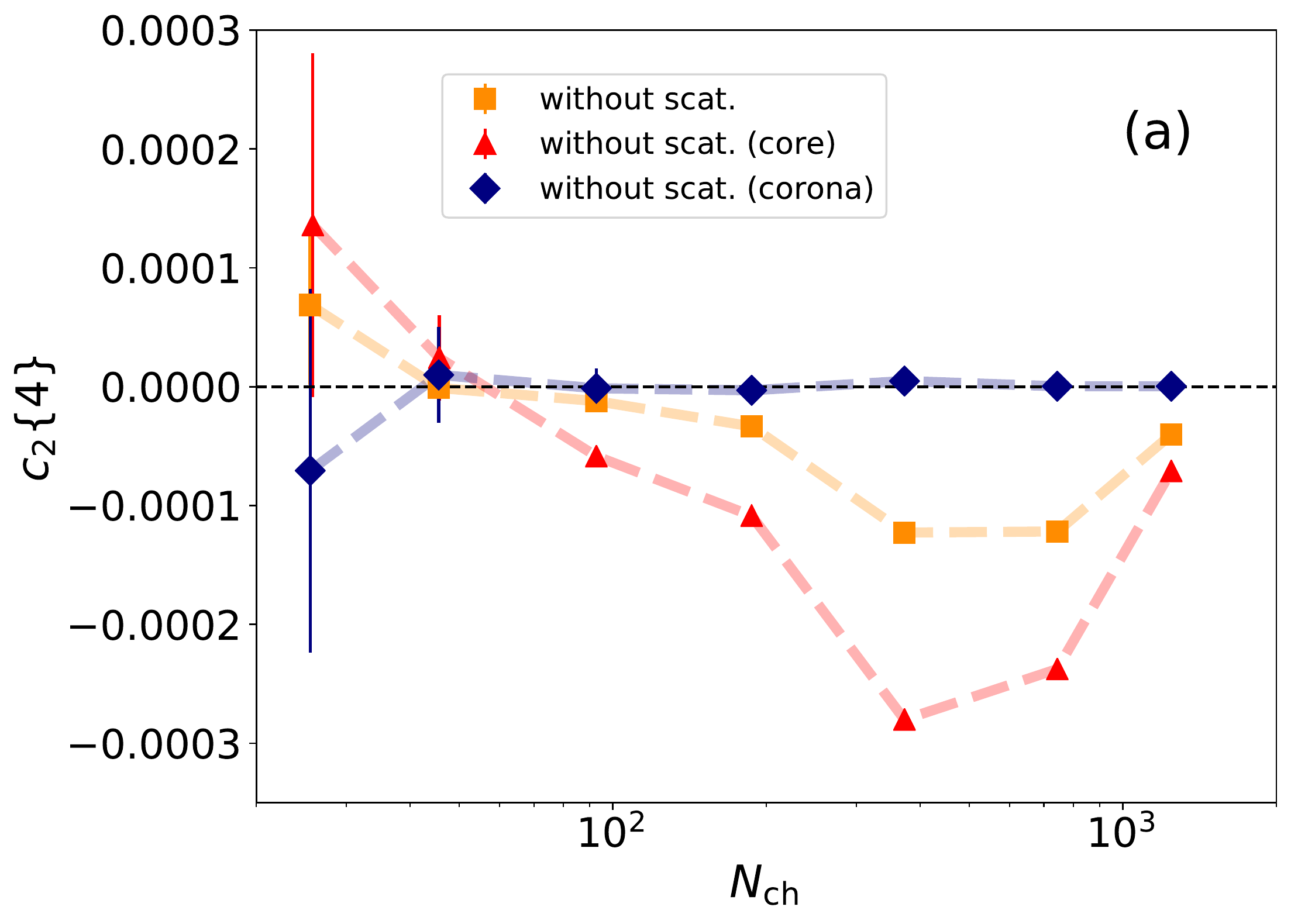}
    \includegraphics[bb=0 0 634 453, width = 0.45\textwidth]{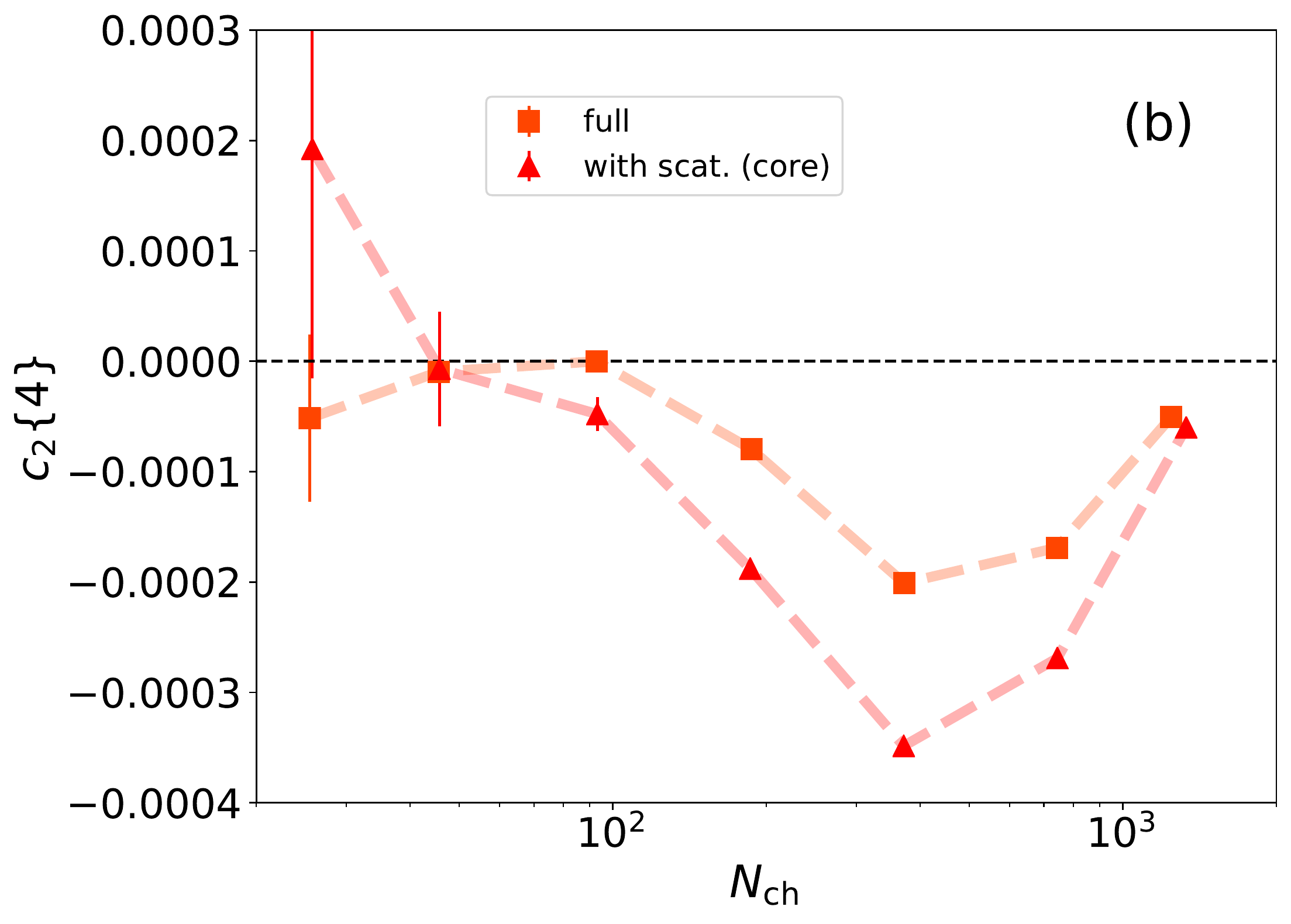}
    \caption{Four-particle cumulants for charged hadrons as functions of the number of produced charged hadrons in Pb+Pb collisions at \snn = 2.76 TeV.
     (a) Results only from the core (red diamonds) and the corona (blue diamonds) components as a breakdown of results without hadronic rescatterings (orange squares) are shown.
     (b) Results from full DCCI2 simulations (squares) and the core components with hadronic rescatterings (triangles) are shown. Dashed lines are drawn as a guide to the eyes.}
    \label{fig:C24_Nch}
\end{figure*}

Figure \ref{fig:PTSPECTRA_LOWPT} shows
$p_{T}$ spectra of charged pions, charged kaons, and protons and antiprotons  in $|\eta|<0.8$ from DCCI2 compared with ALICE experimental data \cite{ALICE:2015dtd} in Pb+Pb collisions at \snn =  2.76 TeV.
In addition to the results from full DCCI2 simulations, the results from the core components with hadronic rescatterings are shown for comparison
in the upper panels of Fig.~\ref{fig:PTSPECTRA_LOWPT}. 
This makes it possible to quantify how much the corona contribution to full DCCI2 simulations compensates for the discrepancy between the experimental data and 
the results from conventional hybrid models.
Shown in the upper panels are the insets to enlarge the very low $p_{T}$ regions in which the corona components are expected to contribute.
In the corresponding lower panels, the ratios of the ALICE experimental data to results with full DCCI2 simulation and to the core components with hadronic rescatterings are shown at each $p_{T}$ bin.

We first observe that the ALICE experimental data cannot be reproduced solely by the core components with hadronic rescatterings at the very low $p_{T}$ region in all cases, which is consistent with previous findings in hydrodynamic models \cite{Nijs:2020ors,Devetak:2019lsk}.
The results from full DCCI2 simulations show better agreement with experimental data than these of the core components with hadronic rescatterings.
Results from full DCCI2 simulations still lack pion yields at the very low $p_{T}$ regions.
On the other hand, the corona contributions fill the discrepancy at the very low $p_{T}$ region between the experimental data and the results from the core components with hadronic rescatterings in the case of kaons and protons.
Therefore one can say that the contribution from the corona components would be one of the strong clues to resolving a long-standing issue of discrepancy between experimental data and hydrodynamic models.


From Figs.~\ref{fig:CORECORONA_PTSPECTRA} and \ref{fig:PTSPECTRA_LOWPT}, conventional hydrodynamic models without the corona contribution reasonably work to describe $p_{T}$-differential observables  in the intermediate $p_{T}$ region ($1 \lesssim p_{T} \lesssim 3$ GeV).
However, $p_{T}$-integrated observables such as mean $p_{T}$ and anisotropic flow coefficients $v_{n}$ can be largely affected by the contribution from the corona components.
To see this, we further investigate the effects of corona contributions on flow observables.
Figures \ref{fig:C24_Nch} (a) and (b) show four-particle cumulants, $\ctwofour$, as functions of charged particle multiplicity, $N_{\mathrm{ch}}$,
in Pb+Pb collisions at \snn = 2.76 TeV.
Here $c_{2}\{4\}$ is decomposed into 4-particle correlation, $\langle \langle 4 \rangle \rangle$, and 2-particle correlation, $\langle \langle 2 \rangle \rangle$:
\begin{align}
     c_{n}\{4\}  & = \langle \langle 4 \rangle \rangle - 2(\langle \langle 2\rangle \rangle)^2, \\
     \label{eq:multi-particle-correlations}
    \langle \langle m \rangle \rangle & = \frac{\sum\limits_{e} 
 \ {_{M_{e}\hspace{-1pt}}P_m}\ \langle m \rangle_{e}}{\sum\limits_{e} \ {_{M_{e}\hspace{-1pt}}P_m} },\\
    \langle 4 \rangle_{e} & = \frac{\sum\limits_{i\neq j\neq k\neq l} e^{in(\phi_i + \phi_j- \phi_k-\phi_l)}}{ _{M_{e}\hspace{-1pt}}P_4 },\\
    \langle 2 \rangle_{e} & = \frac{\sum\limits_{i\neq j}e^{in(\phi_i - \phi_j)}}{ _{M_{e}\hspace{-1pt}}P_2 },
\end{align} 
where $\phi_i$ is an azimuthal angle of the $i$th particle in the $e$th event and $M_{e}$ is the number of charged particles in a kinematic range in that event \cite{Borghini:2000sa}.
Charged particle multiplicity, $N_{\mathrm{ch}}$, is obtained by counting the number of charged particles with $|\eta|<0.8$ and $0.2<p_{T}<3.0$ GeV which is the same kinematic range used in Ref.~\cite{Acharya:2019vdf}.
There is no eta gap imposed in this analysis since we checked that there is no significant difference in results between with and without eta gap in Pb+Pb collisions.
In general, four-particle cumulants, $c_{2}\{4\}$, is negative  when the second order anisotropic flow is finite \cite{Borghini:2000sa}. In fact, we have checked that \pythia8 Angantyr, which does not contain any secondary scattering effects in the default settings, predicts almost zero-consistent $c_{2}\{4\}$ as expected.

In Fig.~\ref{fig:C24_Nch} (a), $\ctwofour$ from simulations without hadronic rescatterings and ones from 
core and corona contributions are shown for comparison.
The results without hadronic rescatterings and the core contributions show negative $\ctwofour$ for above $N_{\mathrm{ch}}\approx 10^2$, which manifests a clear signal of anisotropic collective behavior.
On the other hand, the corona contributions show zero-consistent $\ctwofour$ within statistical error bars for the entire range of $N_{\mathrm{ch}}$.
This clearly demonstrates that multi-particle cumulants are essential indicators for hydrodynamic behaviors of the produced matter.
We point out that the absolute values of $\ctwofour$ solely from the core contributions are diluted to the one without hadronic rescatterings due to the existence of the corona contributions.
As shown in Eq.~(\ref{eq:multi-particle-correlations}), the multi-particle correlation $\langle \langle m \rangle \rangle$ is the correlation per permutation of $m$ particles.
Even if there is a sub-ensemble of particles with no correlations among them which does not contribute to the numerator in Eq.~(\ref{eq:multi-particle-correlations}), 
those particles are counted in the permutation of $m$ particles in the denominator in Eq.~(\ref{eq:multi-particle-correlations}).\footnote{Of course, we are aware that there might exist correlations between the main ensemble (the core) and the sub-ensemble (the corona). We simply assume those correlations do not contribute to the numerator.}
Thus, one can interpret this result as that the correlation originating from the core contributions is 
diluted by the corona contributions 
mainly due to more permutation of particles.
This brings us to the following conclusion that
flow coefficients obtained from conventional hydrodynamic or hybrid models should not be compared with experimental data as long as the non-equilibrium contribution that is manifestly distinguished from the hydrodynamic (core) components exists in the system.

Figure \ref{fig:C24_Nch} (b) shows the comparisons of $\ctwofour$ from full DCCI2 simulations and that from core contributions with hadronic rescatterings.
Compared to the results in Fig.~\ref{fig:C24_Nch} (a), 
one sees that the hadronic rescatterings slightly enhance the absolute values of $\ctwofour$ for both results due possibly to the generation of extra anisotropic flow in the late hadronic stage.
It turns out that there still exists a clear difference between these two results even after rescatterings of hadrons are taken into account in the late hadronic stage.
Therefore the existence of the corona components clearly affects four-particle cumulants as a $p_{T}$-integrated observable.

\textit{Summary.} We analyzed Pb+Pb collisions at \snn = 2.76 TeV from DCCI2
and quantified the corona contributions to particle-identified $p_{T}$ spectra in central and peripheral events.
We found the relative increase of 
the corona contributions below $p_{T} \approx 1$ GeV in both central and peripheral events.
The blue shift of the core spectra, due to the mass of particles, results in a relative increase of corona contributions at very low $p_{T}$ regions.
Especially, we found that the proton and antiproton $p_{T}$ spectra at the central events show $\approx 50\%$ of the corona contributions at very low $p_{T}$ region ($p_{T}\approx 0$ GeV).
From the comparisons between experimental data and results from DCCI2,
we concluded that the corona contributions can be a possible candidate to compensate for the lack of yields obtained from conventional hydrodynamic or hybrid models below $p_{T} \approx 0.5$ GeV.
We also showed that the corona contributions at the very low $p_{T}$ region dilute four-particle cumulants $\ctwofour$ obtained purely from the core contributions.
This result suggests the necessity of the implementation of the corona components to more precisely extract transport coefficients of the QGP from comparisons between dynamical models and experimental data by using Bayesian parameter estimation \cite{Bernhard:2015hxa,Bernhard:2016tnd,Auvinen:2017fjw,Bernhard:2019bmu,JETSCAPE:2020mzn,JETSCAPE:2020shq,Nijs:2020ors,Nijs:2020roc,Auvinen:2020mpc,Parkkila:2021tqq}.

Note that we admit that there is still a slight lack of the very low $p_T$ yields as shown in Fig.~\ref{fig:PTSPECTRA_LOWPT}.
This is highly related to a lack of high $p_{T}$ yields,
which is attributed to the absence of $N_{\mathrm{coll}}$-scaling in \pythia8 Angantyr.
Since the relative increase of the very low $p_{T}$ yields is brought by fragmentation of strings including hard corona partons, it is necessary to improve the description of initial parton production at the high $p_T$ regions in nucleus-nucleus collisions for more quantitative discussions.

\textit{Acknowledgement.} The work by Y.K. is supported by JSPS KAKENHI Grant Number 20J20401. 

\bibliography{inspire.bib}

\end{document}